\documentclass{aastex631}

\def \arcsecond#1#2{$#1''\!\!.#2$}

\begin{document}

\title{Dusty Star Formation Halfway To Cosmic Noon}

\title{Halfway to the peak: Spatially resolved star formation and kinematics in a z=0.54 dusty galaxy with JWST/MIRI}

\author[0000-0002-5830-9233]{Jason Young}
\affiliation{Department of Astronomy, University of Massachusetts, Amherst, MA 01003, USA}

\author[0000-0001-8592-2706]{Alexandra Pope}
\affiliation{Department of Astronomy, University of Massachusetts, Amherst, MA 01003, USA}

\author[0000-0002-1917-1200]{Anna Sajina}
\affiliation{Department of Physics \& Astronomy, Tufts University, Medford, MA, USA}

\author[0000-0003-1710-9339]{Lin Yan}
\affiliation{Division of Physics, Mathematics and Astronomy, California Institute of Technology, Pasadena, CA 91125, USA}

\author[0000-0003-2374-366X]{Thiago S Gon\c{c}alves}
\affiliation{Federal University of Rio de Janeiro, Valongo Observatory, Ladeira Pedro Antonio, 43, Saude 20080-090 Rio de Janeiro, Brazil}

\author[0000-0002-4690-4502]{Miriam Eleazer}
\affiliation{Department of Astronomy, University of Massachusetts, Amherst, MA 01003, USA}


\author[0000-0002-8909-8782]{Stacey Alberts}
\affiliation{Steward Observatory, University of Arizona, 933 N. Cherry Avenue, Tucson, AZ 85721 USA}

\author[0000-0003-3498-2973]{Lee Armus}
\affiliation{IPAC, California Institute of Technology, 1200 East California Boulevard, Pasadena, CA 91125, USA}

\author[0000-0001-9139-2342]{Matteo Bonato}
\affiliation{INAF-Istituto di Radioastronomia and Italian ALMA Regional Centre, Via Gobetti 101, I-40129, Bologna, Italy}

\author[0000-0002-5782-9093]{Daniel A. Dale}
\affiliation{Department of Physics and Astronomy, University of Wyoming, Laramie, WY 82071, USA}

\author[0000-0003-1748-2010]{Duncan~Farrah}
\affiliation{Department of Physics and Astronomy, University of Hawai`i at M\=anoa, 2505 Correa Rd., Honolulu, HI, 96822, US}

\author[0000-0001-6266-0213]{Carl Ferkinhoff}
\affiliation{Department of Physics, Winona State University, Winona, MN, 55987, USA}

\author[0000-0003-4073-3236]{Christopher C. Hayward}
\affiliation{Center for Computational Astrophysics, Flatiron Institute, 162 Fifth Avenue, New York, NY 10010, USA}

\author[0000-0002-6149-8178]{Jed McKinney}
\affiliation{Department of Astronomy, The University of Texas at Austin, Austin, TX, USA}

\author[0000-0001-7089-7325]{Eric J.\,Murphy}
\affiliation{National Radio Astronomy Observatory, 520 Edgemont Road, Charlottesville, VA 22903, USA}

\author[0000-0001-5783-6544]{Nicole Nesvadba}
\affiliation{Université de la Côte d’Azur, Observatoire de la Côte d’Azur, CNRS, Laboratoire Lagrange, Bd de l’Observatoire, CS 34229, 06304 Nice cedex 4, France}

\author[0000-0002-3471-981X]{Patrick Ogle}
\affiliation{Space Telescope Science Institute, 3700 San Martin Drive, Baltimore, MD 21218}

\author[0000-0002-8502-8947]{Leonid Sajkov}
\affiliation{Department of Physics \& Astronomy, Tufts University, Medford, MA, USA}

\author[0000-0002-3158-6820]{Sylvain Veilleux}
\affiliation{Department of Astronomy and Joint Space-Science Institute, University of Maryland, College Park, MD 20742, USA}

\received{August 31, 2023}
\accepted{to ApJL October 9, 2023}

\begin{abstract}

We present JWST/MIRI/MRS observations of an infrared luminous disk galaxy, FLS1, at z$\sim$0.54. With a lookback time of 5~Gyr, FLS1 is chronologically at the midpoint between the peak epoch of star formation and the present day. The MRS data provide maps of the atomic fine structure lines [Ar~II]6.99~\micron{}, [Ar\,III]8.99~\micron{}, [Ne~II]12.81~\micron{}, and [Ne~III]15.55~\micron{}, polycyclic aromatic hydrocarbon (PAH) features at 3.3~\micron{}, 6.2~\micron{}, and 11.3~\micron{}, and the  warm molecular gas indicators H$_2$S(5) and H$_2$S(3); all these emission features are spatially resolved. We find that the PAH emission is more extended along the Northern side of the galaxy when compared to the well-studied star-formation tracer [Ne~II]. The H$_2$ rotational lines, which are shock indicators, are strongest and most extended on the Southern side of the galaxy. {[Ar~II] is the second brightest fine structure line detected in FLS1 and we show that it is a useful kinematic probe which can be detected with JWST out to z$\sim$3.} Velocity maps of [Ar~II]  show a rotating disk with signs of turbulence. Our results provide an example of how spatially resolved mid-infrared spectroscopy can allow us to better understand the star formation and ISM conditions in a galaxy halfway back to the peak epoch of galaxy evolution. 

\end{abstract}


\section{Introduction} \label{sec:intro}

Galaxies have experienced significant evolution during the past 10 billion years with a steady decline in both the star formation and black hole accretion rate densities. Given the pervasive dust obscuration, the energy released by both processes is primarily emitted in the infrared \citep[e.g.,][]{Madau2014}. With the Mid-InfraRed Instrument (MIRI) on JWST, we can spatially resolve multiple spectral tracers of dust-obscured star formation beyond the local Universe to better understand the global evolution of galaxies.

The InfraRed Spectrograph (IRS) on Spitzer provided spectra that could separate star formation and active galactic nuclei (AGN) activity in galaxies out to z$\sim$4 \citep{Yan2007,Pope2008,Riechers2014}, but these observations were spatially unresolved. Spitzer and Infrared Space Observatory (ISO) enabled spatially resolved measurements of the mid-infrared continuum and line emission in nearby galaxies \citep[e.g.,][]{LeFloch2001,Dale2009,DiazSantos2010,DiazSantos2011}.  Several mid-infrared studies suggested that star formation at high redshift was likely more spatially extended than in the local universe \citep[e.g.,][]{Rigby2008,MenendezDelmestre2009}. The Atacama Large Milimeter Array has since resolved the cold dust and gas in z=2-3 sub-milimeter galaxies and found that the cold dust is usually more compact than the stellar mass and the cold molecular gas \citep[e.g.,][]{Lang2019,CalistroRivera2018}. With MIRI Medium Resolution Spectrograph (MRS), we can look at warm star-forming gas to determine how it is distributed in galaxies locally and at high redshift in order to measure any evolution since the peak epoch of star formation.

Given the sensitivity and resolution of MIRI/MRS, we now resolve polycyclic aromatic hydrocarbon (PAH) destruction and ionization as a function of radiation field harshness within individual galaxies \cite[e.g.][]{Lai2022,Armus2023}. Spatial variations in PAH emission have even been resolved in a lensed galaxy at $z=4.2$ \citep{Spilker2023}. The high spectral resolution has also enabled kinematic studies of shocks and outflows \citep[e.g.,][]{U2022,Goold2023,Rich2023}. Lines such as [Ar~II]6.99~\micron{} are strong in nearby infrared-luminous galaxies \citep[e.g.][]{U2022}, and have yet to be explored beyond the local Universe. 

In this letter, we report on new JWST MIRI/MRS observations of FLS1\footnote{Also known as MIPS562 and WISEA J171239.73+585955.1}. These observations allow us, for the first time, to measure the spatial extent of multiple atomic, molecular, and PAH emission features in a dusty star-forming galaxy at z=0.54 \citep{Sajina2012}. FLS1 is drawn from the {\it Halfway to the Peak} sample, which consists of eight massive ($M_* \sim 10^{10}-10^{11} M_\odot$) IR-luminous ($L_{IR} \sim 10^{11}-10^{12}L_\odot$) galaxies with Spitzer IRS redshifts from z$\approx$0.5-0.6. In Hubble Space Telescope (HST) F160W (rest-frame optical) imaging FLS1 appears to be a large disk galaxy with signs of extended emission \citep[][]{Zamojski2011}. From low resolution Spitzer/IRS spectra, FLS1 shows strong star forming lines suggesting a minimal AGN contribution \citep{Sajina2012}. Our detailed analysis of FLS1 allows us to explore the resolved star forming conditions in a galaxy halfway back to the peak epoch of galaxy evolution, a time when the overall rate of star-formation in galaxies was rapidly falling.

In Section~\ref{sec:observations} we describe our observations and data reduction methods. In Section~\ref{sec:results} we discuss the results of spectral extraction and line profile fitting. In Section~\ref{sec:discussion} we examine these results and draw conclusions about the physical nature of FLS1, followed by a summary in Section~\ref{sec:summary}. In this letter, we assume a cosmology of $H_0 = 70~\rm km~s^{-1}~Mpc^{-1}$, $\Omega_M= 0.30$, and $\Omega_\Lambda = 0.70$, {and a \cite{Kroupa2001} IMF.}

\section{Observations and Data Reduction}
\label{sec:observations}

Observations of FLS1 were collected as part of GO program 1762 using the JWST MIRI MRS integral field unit \citep{Wright2023} with 2220$\,$s in each of the three subbands. {This program consisted of eight objects, six of which, including FLS1, are from the same field and were observed within a 13-day span in July/August 2022.} We did not use a dedicated background since the target was predicted to be small enough that we could use the off source pixels for background subtraction. Uncalibrated observations of the source were reduced and assembled into spectral data cubes using the standard JWST Science Calibration Pipeline 1.9.6 \citep{Bushouse2023} with CRDS release \texttt{jwst\_1100.pmap}, with several customizations summarized below. A complete description of our data reduction procedure will be presented in Young et al. (in prep.).

\begin{figure}
\plotone{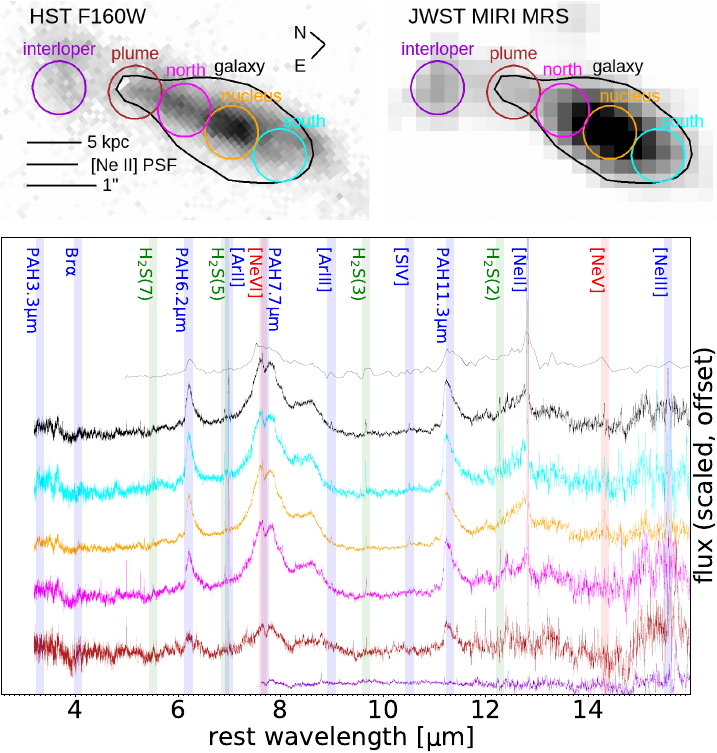}
\caption{{\bf top row: }Collapsed image of all four MRS spectral cubes and HST F160W (rest frame optical) image, with a $4\sigma$ MRS detection contour (full galaxy aperture, black) and five selected regions (colored circles). {\bf bottom: } Spectra extracted from the full galaxy aperture and the five sub-regions, and the archival IRS spectrum (top thin spectrum). {The IRS slit is close to the size of the entire MRS FOV.} Spectra are normalized and offset for visual clarity, and color-coded the same as  the regions. {The vertical shaded bands mark key spectral features.}
\label{fig:jwst_hst}}
\end{figure}

\begin{figure}
\plotone{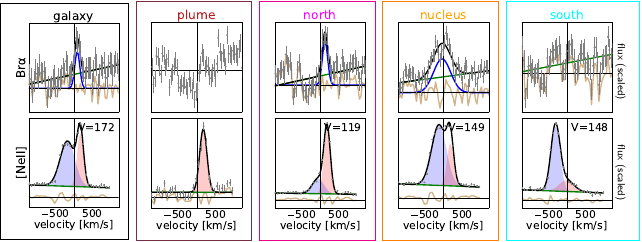}
\caption{Cutouts of the extracted spectrum around Br$\rm\alpha$ (top) and [Ne~II] (bottom), with best-fit Gaussian models. For [Ne~II], red and blue shifted components are shown, and one half of the peak-to-peak velocity differences are indicated. The tan curve at the bottom is the residual between the data and the continuum+Gaussian(s) model.
\label{fig:line_profiles}}
\end{figure}

\begin{enumerate}
    \item  The dark subtraction step in stage~1 was turned off, since the standard dark was not taken with the same number of groups as our data. Calibration data with different number of groups are known to introduce additional noise, which our experiments confirmed.
  
    \item The stage~1 reduction (\texttt{Detector1Pipeline}) was run twice, once with parameters optimized to detect small cosmic rays and again with parameters to detect and repair larger cosmic ray ``showers''.

    \item All six objects observed in July/August were reduced through the standard stage~2 \texttt{Spec2Pipeline} and stage~3 \texttt{Spec3Pipeline} steps to create preliminary cubes for each channel. These cubes were collapsed (averaged over the wavelength axis), and 8$\sigma$ detection contours were generated around each target. These contours were used for masking in the following steps.
    
    \item With the primary targets masked, the stage~1 rate files for the six objects observed in July/August were median combined on a per detector/grating setting basis to generate master backgrounds, which were subtracted from each of the rate files prior to re-running the stage~2 \texttt{Spec2Pipeline} reduction.

    \item Residual fringe corrections were made to the 2D calibrated rate files as a final step within stage~2 reduction \texttt{Spec2Pipeline} using the \texttt{jwst.residual\_fringe} package.

    \item Prior to the stage~3 reduction \texttt{Spec3Pipeline}, residual background subtraction was performed on a per-slice basis using the median of non-masked pixels within each row of each slice. (the ``slices'' are the dispersed spectra from each section of the image slicer).

    \item For analyses requiring 1D spectra extracted from the cubes, the residual fringe removal code \texttt{rfc1d\_utils} was applied to the 1D spectra.

\end{enumerate}

The background removed in steps~4 and 6 includes both astrophysical and instrumental background. In many of the following analyses we compare features at different wavelengths. Unless otherwise stated, these analyses use cubes with all the wavelength slices convolved to the spatial resolution of [Ne~II]12.81~\micron{}. The angular resolution in each wavelength slice was estimated from the in-flight data presented in \cite{Argyriou2023}. In addition to the MRS data, our program also took simultaneous observations with the MIRI imager at 5.6~\micron{}, spatially offset from our primary target. In this letter, we use these parallel imaging data for the purposes of verifying astrometry.

\subsection{Astrometric Corrections}

Figure~\ref{fig:jwst_hst} shows both our collapsed JWST MIRI spectral cube as well as a HST F160W (rest-frame optical) image of this galaxy. The difference in the positions of two stars in the HST image and three stars in the MIRI 5.6~\micron{} image (not shown) to their Gaia positions show offsets of (+\arcsecond{0}{1},-\arcsecond{0}{5}) in the HST image and (+\arcsecond{0}{6},+\arcsecond{0}{3}) in the MIRI image. The HST astrometry was adjusted by these amounts to bring it into alignment with the MIRI data. The Gaia corrected position of the nucleus (orange circle in Figure~\ref{fig:jwst_hst}) is 17:12:39.4 +58:59:54.1.

\subsection{Spectral Extraction}

Figure~\ref{fig:jwst_hst} shows several circular apertures which were used to extract spectra at different positions across the galaxy. Each aperture has a diameter of \arcsecond{0}{75} (4.8~kpc), the in-flight PSF FWHM of MRS at the observed wavelength of [Ne\,II] at z=0.54 \citep{Argyriou2023}. The [Ne\,II] beam was used as an aperture because it is the longest wavelength used in most of our analyses. The larger irregular black line denotes a 4$\sigma$ MRS detection contour from which we extracted the total galaxy spectrum. The lower panel in Figure\,\ref{fig:jwst_hst} shows the extracted spectra from each of these apertures, using the same color codes. The interloper has a spectrum that is not consistent with the redshift of FLS1, so we do not consider it in this letter.

Our analysis that uses the smaller beam-sized apertures is focused on comparisons of line flux ratios, and so no aperture corrections are needed. We have also made no attempt to correct for extinction. Even in dusty galaxies, mid-IR extinction is expected to be mild and relatively flat \citep{Stone2022},  except in silicate and ice absorption bands \citep{Draine2003}. The spectral features discussed in this letter are mostly outside those bands, with the exception of H$_2$S(3).

\subsection{Kinematic analysis}

Due to its high signal-to-noise, isolation from other atomic lines, and high spatial resolution, we use the [Ar\,II] line to perform a kinematic analysis. We follow the procedure described in \cite{Goncalves2010}. In summary, we fit a single Gaussian profile to each spaxel, limited to $\pm 1000$ km s$^{-1}$ of the center of the integrated emission, with an additional free parameter to allow an average continuum value underneath the line. The position of the peak of the Gaussian profile determines the velocity of that spaxel, while the standard deviation $\sigma$ of the Gaussian fit is adopted as the velocity dispersion. The signal-to-noise is determined from the standard deviation of spaxels away from the galaxy, where only background noise is expected. The angular resolution in our velocity maps is \arcsecond{0}{426} (2.7~kpc at z=0.54), the spatial resolution at the observed wavelength of [Ar\,II].

\section{Results}
\label{sec:results}

\subsection{Line strengths and profiles}
\label{sec:lineprofiles}

Figure~\ref{fig:jwst_hst} marks the key emission features in each extracted spectrum.  The PAH 6.2, 7.7, and 11.3~$\rm\mu$m features and the atomic star-formation indicators [Ne\,II], [Ar\,II], [Ne\,III]15.55~\micron{}, and [Ar\,III]8.99~\micron{} are visible in all of the regions of FLS1. [Ne\,II] and  [Ar\,II] are particularly bright. Higher ionization lines indicative of AGN, [Ne~V]14.32~$\rm\mu$m and [Ne VI]7.65~$\rm\mu$m, are visible in the nucleus. We also see strong detections of the molecular hydrogen lines H$_2$S(3) and H$_2$S(5), and weaker detections of H$_2$S(2), H$_2$S(4), and H$_2$S(7). For comparison, the top thin spectrum is the archival IRS spectrum for FLS1. While the broad characteristics of the spectrum and the brighter PAH bands are visible in the IRS spectrum, most of the other lines are either undetected or unresolved. We verify the mid-infrared AGN fraction from the IRS spectrum by performing the exact same analysis on the new MRS full galaxy spectrum. Following the procedure in \cite{Pope2008}, we fit a simple model consisting of a star formation template, power law emission from an AGN and extinction \citep{Draine2003}. We find a mid-IR AGN fraction of $10\pm2\%$ which suggests that the AGN does not have a significant effect on the overall energetics in the mid-IR. In this letter, we focus on the spatially resolved lines tracing star formation and shocked gas, and the AGN lines will be explored in a future paper.

A key advantage of the MRS data is that most of our features are spectrally resolved. 
Figure~\ref{fig:line_profiles} shows Gaussian line profile fits to Br$\rm\alpha$ and [Ne\,II]. 
Based on the in-flight performance analysis \citep{Argyriou2023}, the MRS data have a velocity resolution of 79~km/s and 144~km/s at Br$\rm\alpha$ and [Ne\,II], respectively. Based on the fit to the whole-galaxy Br$\rm\alpha$, we find a redshift of 0.5385$\pm$0.0007, which we adopt as our systemic redshift. The galaxy and nucleus [Ne\,II] profiles clearly show double-peaked structure, so we fit the [Ne\,II] profiles with a linear continuum plus a double Gaussian model. In all cases, the velocity half separations between components were found to be less than 200~km/s. Br$\rm\alpha$ was detected in the north, nucleus, and galaxy regions; the line shows a single peak in kinematic alignment with the bluer [Ne\,II] peaks. 

Similar fits were performed for [Ne\,II], [Ar\,II] and [Ar\,III] and the strongest molecular hydrogen features, H$_2$S(5) and H$_2$S(3). Table~\ref{tab:emission_line_luminosities} lists the integrated line fluxes. The  uncertainties are the standard deviations of the flux values from 1,000 Monte-Carlo realizations in which Gaussian noise corresponding to the measured 1\,$\sigma$ uncertainty was added to each spectral point and the fitting was repeated.

\subsection{Star Formation Rates}
\label{sec:sfr}

{We compare different MIR SFR indicators using the spectral lines from Table~1, which have not been corrected for extinction. Using the [Ne~II] measurement for the full galaxy and the empirical LIR/[Ne\,II] relationship in \cite{Ho2007} we estimate  log(LIR/L$_\odot$) of 11.93, in close agreement with the Spitzer+Herschel derived 11.91 cited in \cite{Sajina2012}. With the SFR/LIR relationship from \cite{Murphy2011}, we calculate a [Ne\,II] galaxy-wide SFR of 128~M$_\odot$~yr$^{-1}$ with an uncertainty of 0.5~dex from the [Ne~II]/LIR calibration. Of the total [Ne~II] SFR, 43$\pm$2\% comes from the nucleus.

In order to compare SFRs from other indicators, we focus on the nucleus, where the Br$\rm\alpha$ and [Ne~III] detections are the strongest. The [Ne~II]-derived SFR in the nucleus is 55$\pm$2~M$_\odot$~yr$^{-1}$. Using the [Ne~II]+[Ne~III] SFR relation \citep[][0.26~dex scatter]{Ho2007} we calculate 27$\pm$1 M$_\odot$~yr$^{-1}$, which is consistent with the metallicity dependent relation from  \cite{Zhuang2019} assuming solar metallicity. Since the [Ne~II] SFR from \cite{Murphy2011} assumes a Kroupa IMF and the [Ne~II]+[Ne~III] calibrations assume a Salpeter IMF, we have adjusted the [Ne~II]+[Ne~III] SFR by a factor of 0.68 \citep{Kennicutt2012}. The nucleus Br$\rm\alpha$ emission translates to an SFR of 13$\pm$2~M$_\odot$~yr$^{-1}$, assuming case-B emissivity coefficients for $T_e=10^4~\rm K$ and $n_e=10^2~\rm cm^{-3}$  \citep{Storey1995} and the H$\rm\alpha$/SFR relationship in \cite{Kennicutt2012}. While the uncertainties on the SFR calibrations account for these different SFR values, a number of physical mechanisms could contribute towards the lower Br$\rm\alpha$ SFR, such as differential extinction, the spatial distributions of different tracers, star formation timescales, and continuum absorption. 
 Given the low SNR detection of Br$\rm\alpha$ even in the nucleus, we caution against over-interpreting the Br$\rm\alpha$ SFR of this one source.

Although [Ar~II] is a bright low-ionization line, we did not find any empirical [Ar~II] SFR calibrations in the literature.} Likewise, although the PAH features are strongly detected, they are blended with other features, including the silicate absorption, and a simple extraction assuming a linear continuum is not reliable. {We will explore SFRs from Br$\rm\alpha$, [Ar~II] and PAHs with more sophisticated analysis of our full sample in a future paper.}

\begin{deluxetable}{rccccccc cccc}
\tabletypesize{\footnotesize}
\tablecolumns{2} 
\tablecaption{\label{tab:emission_line_luminosities}Emission Line Fluxes}
\tablehead{ & 
\colhead{Br$\rm\alpha$}\hspace{.2cm} & 
\colhead{[Ne~II]} &
\colhead{[Ne~III]} &
\colhead{[Ar~II]} &
\colhead{[Ar~III]} &
\colhead{H$_2$S(5)} &
\colhead{H$_2$S(3)} \\
&
\colhead{4.05~\micron{}}\hspace{.2cm} & 
\colhead{12.81~\micron{}} &
\colhead{15.56~\micron{}} &
\colhead{6.99~\micron{}} &
\colhead{8.99~\micron{}} &
\colhead{6.91~\micron{}} &
\colhead{9.67~\micron{}} 
} 
\startdata
galaxy	&	103.3	$\pm$	32.4	&	2050.7	$\pm$	36.4	&	340.1	$\pm$	82.3	&	999.9	$\pm$	40.5	&	135.9	$\pm$	15.3	&	199.6	$\pm$	44.3	&	212.1	$\pm$	20.9	\\
north	&	22.7	$\pm$	6.3	&	284.0	$\pm$	3.6	&	29.4	$\pm$	13.3	&	111.2	$\pm$	6.1	&	14.8	$\pm$	4.3	&	22.5	$\pm$	2.3	&	21.6	$\pm$	2.6	\\
nucleus	&	88.1	$\pm$	14.1	&	882.2	$\pm$	27.0	&	228.5	$\pm$	32.6	&	435.4	$\pm$	4.7	&	44.9	$\pm$	2.7	&	82.3	$\pm$	8.0	&	62.1	$\pm$	2.9	\\
plume	&	$<$37.2			&	70.0	$\pm$	4.1	&	$<$21.3			&	21.4	$\pm$	4.1	&	$<$16.7			&	5.9	$\pm$	1.9	&	$<$14.0			\\
south	&	$<$38.8			&	279.4	$\pm$	9.2	&	$<$22.3			&	139.2	$\pm$	2.8	&	18.9	$\pm$	3.2	&	35.5	$\pm$	2.5	&	37.5	$\pm$	2.2	\\
\label{tbl:atomic_lines}
\enddata 
\tablenotetext{\tiny \dagger}{The line fluxes are in units of $10^{-21}\rm W\,m^{-2}$, and have not been corrected for extinction.}
\tablenotetext{\tiny \dagger\tiny \dagger}{~The listed uncertainties are 1$\sigma$.}
\tablenotetext{\tiny \dagger\tiny \dagger\tiny \dagger}{~~3$\sigma$ upper limits are given when values are below 3$\sigma$.}

\end{deluxetable}

\begin{figure}
\plotone{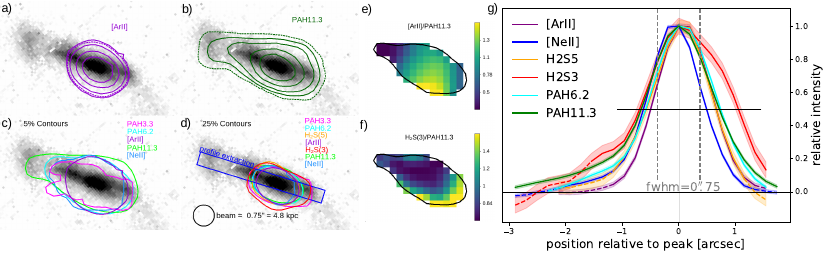}
\caption{
 {\bf a,b)} 
Contours generated from the narrow-band images for [Ar~II] and PAH11.3 (convolved to the spatial resolution of [Ne~II]12.81~\micron{}) overlaid on the HST~F160W image. The dashed contour indicates the 2$\sigma$ detection threshold, as determined by the RMS noise in background areas of the narrow-band images. The solid contours are 50\%, 25\%, 10\%, and 5\% of the peak value in the narrow-band images. {\bf c)}  
 5\% contours for select emission features. {\bf d)} 25\% peak levels for select emission features. All contours are generated from the narrow-band images convolved to the spatial resolution of [Ne~II], except the 5\% PAH~3.3~\micron{} contours, where the faintness of the spectrum meant that convolved image was dominated by noise.  {\bf e,f)} Ratio maps of the [Ar\,II] and H$_2$S(3) lines to PAH11.3 derived from the narrow-band images, normalized to their respective average values. {\bf g)} Normalized spatial profiles extracted from the rectangular aperture in panel d. The contours become dashed when they are below the 2$\sigma$ threshold.  All the emission features are more extended than the [Ne~II]12.81~\micron{} beam (vertical dashed lines). 
\label{fig:jwst_hst_contours}}
\end{figure}

\subsection{Spatial Extent of Emission Features}
\label{sec:starformationoffsets}

Figure~\ref{fig:jwst_hst_contours} (panels a-d)  overlays the HST rest-frame optical image with contours generated from narrow-band images of the spectral lines, and panel g compares the spatial emission profiles of those features along the major axis of FLS1. For the atomic and molecular features, the continuum was removed from the narrow-band images by fitting a linear continuum anchored in two flanking windows. Because the PAH features span a much wider spectral range and are blended with other features (such as silicate absorption), the assumption of a linear continuum is less reliable for PAHs. However, PAH features also contain much more flux, making the continuum fit less significant for determining the spatial extent. We found that including continuum subtraction for the PAH features had almost no impact on most of the contours. The exception was the 3.3~\micron{} PAH feature in Figure~\ref{fig:jwst_hst_contours}, where skipping the continuum removal significantly reduced the noise since the continuum is low compared to the noise.

Both the contour plots (panels a-d) and the extracted spatial profiles (panel g) in Figure~\ref{fig:jwst_hst_contours} show that all the features are broader than the [Ne\,II] beam, making FLS1 the first galaxy with resolved detections of the atomic, warm molecular gas, and PAH lines outside the local universe. 
We also find that the PAH spatial profiles are more extended than the atomic features and the warm molecular gas features (considering the same percentile contours, panels e and f). We discuss the implications of the different spatial distributions in Section\,\ref{sec:discussion}.

\begin{figure}
\plotone{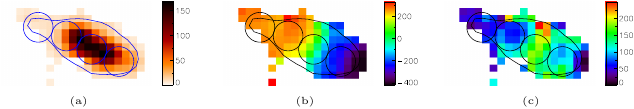}
\caption{ Velocity maps generated by fitting a single Gaussian to the [Ar~II]6.99~\micron{} line in each spaxel. Regions are the same as Figure~\ref{fig:jwst_hst}. {\bf (a)} S/N maps, values are ratios of integrated signal in the fitted line to the quadrature sum of the uncertainties of the spectral elements used. {\bf (b) } Velocity centroid in km/s relative to the adopted systemic velocity. {\bf (c) } Dispersion of the fitted line in km/s.\label{fig:velocity}}
\end{figure}

\subsection{Velocity Field}
\label{sec:velocity}

Figure~\ref{fig:line_profiles} shows that different regions in the disk of FLS1 are offset in velocity. In Figure~\ref{fig:velocity} we show the moment maps of FLS1 for the [Ar\,II] line. The velocity map is consistent with a rotating disk, with no {evidence for large scale systematic departures from circular rotation such as an ongoing major merger}. The in-flight performance analysis \citep{Argyriou2023} indicate a velocity resolution of 92\,km/s at the observed wavelength of [Ar~II], so we cannot rule out minor departures from circular rotation.

We estimate $V_{flat}=\frac{1}{2}\left(V_{max}-V_{min}\right)$, where $V_{min}$ and $V_{max}$ are the medians of the lower and upper 10th percentile velocities  \citep{Goncalves2010}. This gives us $V_{\rm flat} = 317$ km s$^{-1}$. The light weighted average of the line width is 118~km~s$^{-1}$ after correcting for the instrumental width in quadrature. Adopting this line width as a characteristic dispersion $\sigma$ and adopting $V_{flat}$ as a characteristic velocity,  we find $V/\sigma=2.7$, {consistent with a turbulent rotating disk.}

{We investigate whether the kinematics found for [Ar~II] are consistent for other lines. Specifically, we rerun our analysis on [Ne~II], the brightest line in our spectra, and H$_2$S(5), which is in the same channel as [Ar~II]. We find the velocity fields of [Ne~II] and H$_2$S(5) also show rotation with $V/\sigma\sim3$, but have lower spatial resolution and lower signal to noise, respectively.}

\section{Discussion}
\label{sec:discussion}

\begin{figure}
\plotone{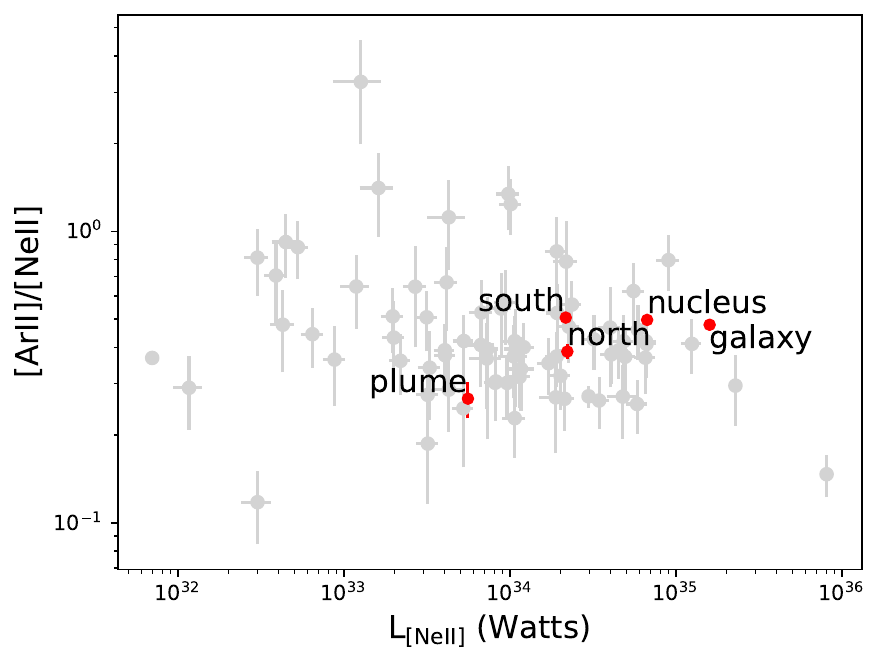}
\caption{[Ar\,II]/[Ne\,II] as a function of [Ne\,II] luminosity for the five regions in FLS1 (red) compared to local star-forming galaxies (grey). FLS1 is broadly in-line with typical star forming galaxies, although most of the points falls on the upper envelope of local [Ar\,II]/[Ne\,II] ratios.  The comparison objects are drawn from the limited and heterogeneous sample of objects with published JWST, Spitzer IRS, and ISO measurements for the necessary lines. In most of the comparison galaxies, the spectral aperture includes only the central region of the object. Since this heterogeneous literature sample has not been corrected for extinction, we have plotted our uncorrected line ratios. Using the optical depths we found in Section~\ref{sec:sfr} would result in a 20\% decrease in the [Ar\,II]/[Ne\,II] ratio. Measurements are taken from \cite{Forster2001,Egami2006,Armus2006,Verma2003,Gallimore2010,Sales2010,Goold2023,Armus2023}; {\cite{Gallimore2010,Sales2010}}.}
\label{fig:line_ratios}
\end{figure}

{Since [Ar~II] is a bright light at a much shorter wavelength than [Ne~II] or [Ne~III], it is well positioned to be a key tracer of star formation and kinematics at higher redshifts in the JWST era. In Figure~\ref{fig:line_ratios} we compare the [Ar\,II]/[Ne\,II] ratio to [Ne\,II] luminosity for all five regions of FLS1 and a range of local galaxies.  The comparison galaxy observations are from Spitzer IRS and ISO Short Wavelength Spectrometer (SWS), with heterogeneous spatial coverage ranging from whole-galaxy measurements to nuclear measurements, and are included to provide a context for the range of values seen in the local universe.  The local measurement span an order of magnitude in [Ar~II]/[Ne~II] ratio, and all five regions of FLS are consistent with the average. This supports the idea that [Ar~II] and [Ne~II] trace similar physical conditions, and shows that [Ar~II] has the potential to trace star formation and kinematics at redshifts where [Ne~II] is inaccessible.}


{The differences in the spatial extent of the emission features in Section 3.3 can be explained by differences in the current and recent star formation, the effect of the AGN, and/or the presence of shocks.} Similar offsets between PAH and atomic lines have been reported in nearby dusty galaxies \citep{DiazSantos2011,Lai2023}. [Ne~II] and [Ar~II], which are sensitive to star formation within the past 10~Myr, are the most compact, while the PAH emission is more extended. PAH emission traces star formation within the past several 100~Myr \citep[e.g.,][]{Smercina2018}. The PAH 11.3~\micron{} emission, which traces neutral PAHs, is the most extended feature we detect in FLS1. The PAH~6.2~\micron{} emission, which is more sensitive to ionized PAHs \citep{Draine2021}, is intermediate in spatial extent between the PAH~11.3~\micron{} and [Ne~II] emission. These variations in the spatial extent of the PAH and atomic features are consistent with an ionization gradient driven by current star formation in the nucleus compared to recent star formation in the disk. We cannot rule out the influence of the AGN on the nuclear emission, in which case some degree of the ionization gradient could be attributed to the AGN.

The diminished spatial extent of the [Ne~II], [Ar~II], and PAH emission in the south compared to the north could be explained by a more turbulent environment in the southern disk that is less favorable for star formation. This scenario is supported by Figure~\ref{fig:velocity}c, where we see that the [Ar\,II] line width in the south region is around 100~km/s, compared to the 50~km/s typical of the north and plume regions. The shock-tracing H$_2$S(3) line extends into the southern disk (Figure~\ref{fig:jwst_hst_contours}f,g), unlike fine structure and PAH features, further supporting the picture of a more disturbed environment in the southern disk. Given the high star-formation rate of FLS1, a recent gas accretion event may be the cause of the turbulence seen in the southern disk. In addition to shocks, warm H$_2$ gas can also be heated by UV photons from young stars or by x-rays; however, the high H$_2$/PAH contrast seen in the southern disk makes star formation an unlikely power source \citep[e.g.,][]{Nesvadba2010,U2023}, and the low AGN fraction makes x-rays unlikely.

The overall picture that emerges is consistent with current star formation in the nucleus, recent star formation in the northern disk, and shocks in the southern disk, although these data admit other interpretations.  In Section~\ref{sec:lineprofiles} we found that the [Ne\,II] profiles are either single or double peaked, and in Section~\ref{sec:velocity}  we found that  $V/\sigma=2.7$. Therefore, with no kinematic signs of disturbance, FLS1 is not likely undergoing a major merger. However, the hypothesized southern shocks and central starburst are in-line with a recent gas accretion event.

\section{Summary}
\label{sec:summary}

This work highlights the rich assortment of mid-IR spectral diagnostics that are spatially resolved in distant galaxies with MIRI MRS. It demonstrates the power of high spectral resolution spectroscopy to identify rotating disks, even at high redshift. FLS1 and the other galaxies in the {\it Halfway to the Peak} program represent a key stepping stone between cosmic noon and the modern universe. Our findings can be summarized as follows:

\begin{enumerate}
  
  \item {While the Br$\rm\alpha$ SFR is roughly half that from the neon lines, they are consistent given the lower SNR detection of Br$\rm\alpha$ and scatter in SFR calibrations.}
  
  \item We detect spatially extended emission from the star-formation tracers [Ne~II] and [Ar~II], PAH features at 3.3~, 6.2 and 11.3~\micron{}, and the warm molecular gas tracers H$_2$S(5) and H$_2$S(3). 
  
  \item PAH~11.3~\micron{} emission is extended on the northern side of FLS1 and not detected on the southern side. All the PAH, atomic, and molecular features discussed are brightest in the nucleus. These observations are consistent with current star formation in the nucleus, recent star formation in the north, and shocked gas in the south.
  
  \item The kinematics show a rotating disk with $V/\sigma=2.7$, suggesting that FLS1 is not the site of a major merger.

  \item {The [Ar\,II]/[Ne\,II] ratios in FLS1 are consistent with those seen in local galaxies.}

\end{enumerate}

{In a forthcoming paper we will expand upon these analyses with a full spectral decomposition of the entire sample, which will allow us to test these findings across all eight objects.}

\section*{Acknowledgments}

We thank the anonymous reviewer for their thoughtful comments, which we believe have improved our manuscript. We are grateful to the following individuals who helped with the analysis of these data: David Law, Jane Morrison, Dick Shaw, Bryan Holler, Beth Sargent, and Sean Linden. Based on observations with the NASA/ESA/CSA James Webb Space Telescope obtained at the Space Telescope Science Institute, which is operated by the Association of Universities for Research in Astronomy, Incorporated, under NASA contract NAS5-03127. Support for program number JWST-GO-01762 was provided through a grant from the STScI under NASA contract NAS5-03127.
The Flatiron Institute is supported by the Simons Foundation.

\bibliography{halfway2peak}{}

\begin{thebibliography}{}
\expandafter\ifx\csname natexlab\endcsname\relax\def\natexlab#1{#1}\fi
\providecommand{\url}[1]{\href{#1}{#1}}
\providecommand{\dodoi}[1]{doi:~\href{http://doi.org/#1}{\nolinkurl{#1}}}
\providecommand{\doeprint}[1]{\href{http://ascl.net/#1}{\nolinkurl{http://ascl.net/#1}}}
\providecommand{\doarXiv}[1]{\href{https://arxiv.org/abs/#1}{\nolinkurl{https://arxiv.org/abs/#1}}}

\bibitem[{{Argyriou} {et~al.}(2023){Argyriou}, {Glasse}, {Law}, {Labiano},
  {{\'A}lvarez-M{\'a}rquez}, {Patapis}, {Kavanagh}, {Gasman}, {Mueller},
  {Larson}, {Vandenbussche}, {Glauser}, {Royer}, {Dicken}, {Harkett},
  {Sargent}, {Engesser}, {Jones}, {Kendrew}, {Noriega-Crespo}, {Brandl},
  {Rieke}, {Wright}, {Lee}, \& {Wells}}]{Argyriou2023}
{Argyriou}, I., {Glasse}, A., {Law}, D.~R., {et~al.} 2023, arXiv e-prints,
  arXiv:2303.13469, \dodoi{10.48550/arXiv.2303.13469}

\bibitem[{{Armus} {et~al.}(2006){Armus}, {Bernard-Salas}, {Spoon}, {Marshall},
  {Charmandaris}, {Higdon}, {Desai}, {Hao}, {Teplitz}, {Devost}, {Brandl},
  {Soifer}, \& {Houck}}]{Armus2006}
{Armus}, L., {Bernard-Salas}, J., {Spoon}, H.~W.~W., {et~al.} 2006, \apj, 640,
  204, \dodoi{10.1086/500040}

\bibitem[{{Armus} {et~al.}(2023){Armus}, {Lai}, {U}, {Larson}, {Diaz-Santos},
  {Evans}, {Malkan}, {Rich}, {Medling}, {Law}, {Inami}, {Muller-Sanchez},
  {Charmandaris}, {van der Werf}, {Stierwalt}, {Linden}, {Privon},
  {Barcos-Mu{\~n}oz}, {Hayward}, {Song}, {Appleton}, {Aalto}, {Bohn},
  {B{\"o}ker}, {Brown}, {Finnerty}, {Howell}, {Iwasawa}, {Kemper}, {Marshall},
  {Mazzarella}, {McKinney}, {Murphy}, {Sanders}, \& {Surace}}]{Armus2023}
{Armus}, L., {Lai}, T., {U}, V., {et~al.} 2023, \apjl, 942, L37,
  \dodoi{10.3847/2041-8213/acac66}

\bibitem[{{Bushouse} {et~al.}(2023){Bushouse}, {Eisenhamer}, {Dencheva},
  {Davies}, {Greenfield}, {Morrison}, {Hodge}, {Simon}, {Grumm}, {Droettboom},
  {Slavich}, {Sosey}, {Pauly}, {Miller}, {Jedrzejewski}, {Hack}, {Davis},
  {Crawford}, {Law}, {Gordon}, {Regan}, {Cara}, {MacDonald}, {Bradley},
  {Shanahan}, {Jamieson}, {Teodoro}, \& {Williams}}]{Bushouse2023}
{Bushouse}, H., {Eisenhamer}, J., {Dencheva}, N., {et~al.} 2023, {JWST
  Calibration Pipeline}, 1.9.6, Zenodo,  Zenodo, \dodoi{10.5281/zenodo.7714020}

\bibitem[{{Calistro Rivera} {et~al.}(2018){Calistro Rivera}, {Hodge}, {Smail},
  {Swinbank}, {Weiss}, {Wardlow}, {Walter}, {Rybak}, {Chen}, {Brandt},
  {Coppin}, {da Cunha}, {Dannerbauer}, {Greve}, {Karim}, {Knudsen},
  {Schinnerer}, {Simpson}, {Venemans}, \& {van der Werf}}]{CalistroRivera2018}
{Calistro Rivera}, G., {Hodge}, J.~A., {Smail}, I., {et~al.} 2018, \apj, 863,
  56, \dodoi{10.3847/1538-4357/aacffa}

\bibitem[{{Dale} {et~al.}(2009){Dale}, {Smith}, {Schlawin}, {Armus},
  {Buckalew}, {Cohen}, {Helou}, {Jarrett}, {Johnson}, {Moustakas}, {Murphy},
  {Roussel}, {Sheth}, {Staudaher}, {Bot}, {Calzetti}, {Engelbracht}, {Gordon},
  {Hollenbach}, {Kennicutt}, \& {Malhotra}}]{Dale2009}
{Dale}, D.~A., {Smith}, J.~D.~T., {Schlawin}, E.~A., {et~al.} 2009, \apj, 693,
  1821, \dodoi{10.1088/0004-637X/693/2/1821}

\bibitem[{{D{\'\i}az-Santos} {et~al.}(2010){D{\'\i}az-Santos},
  {Alonso-Herrero}, {Colina}, {Packham}, {Levenson}, {Pereira-Santaella},
  {Roche}, \& {Telesco}}]{DiazSantos2010}
{D{\'\i}az-Santos}, T., {Alonso-Herrero}, A., {Colina}, L., {et~al.} 2010,
  \apj, 711, 328, \dodoi{10.1088/0004-637X/711/1/328}

\bibitem[{{D{\'\i}az-Santos} {et~al.}(2011){D{\'\i}az-Santos}, {Charmandaris},
  {Armus}, {Stierwalt}, {Haan}, {Mazzarella}, {Howell}, {Veilleux}, {Murphy},
  {Petric}, {Appleton}, {Evans}, {Sanders}, \& {Surace}}]{DiazSantos2011}
{D{\'\i}az-Santos}, T., {Charmandaris}, V., {Armus}, L., {et~al.} 2011, \apj,
  741, 32, \dodoi{10.1088/0004-637X/741/1/32}

\bibitem[{{Draine}(2003)}]{Draine2003}
{Draine}, B.~T. 2003, \araa, 41, 241,
  \dodoi{10.1146/annurev.astro.41.011802.094840}

\bibitem[{{Draine} {et~al.}(2021){Draine}, {Li}, {Hensley}, {Hunt},
  {Sandstrom}, \& {Smith}}]{Draine2021}
{Draine}, B.~T., {Li}, A., {Hensley}, B.~S., {et~al.} 2021, \apj, 917, 3,
  \dodoi{10.3847/1538-4357/abff51}

\bibitem[{{Egami} {et~al.}(2006){Egami}, {Rieke}, {Fadda}, \&
  {Hines}}]{Egami2006}
{Egami}, E., {Rieke}, G.~H., {Fadda}, D., \& {Hines}, D.~C. 2006, \apjl, 652,
  L21, \dodoi{10.1086/509886}

\bibitem[{{F{\"o}rster Schreiber} {et~al.}(2001){F{\"o}rster Schreiber},
  {Genzel}, {Lutz}, {Kunze}, \& {Sternberg}}]{Forster2001}
{F{\"o}rster Schreiber}, N.~M., {Genzel}, R., {Lutz}, D., {Kunze}, D., \&
  {Sternberg}, A. 2001, \apj, 552, 544, \dodoi{10.1086/320546}

\bibitem[{{Gallimore} {et~al.}(2010){Gallimore}, {Yzaguirre}, {Jakoboski},
  {Stevenosky}, {Axon}, {Baum}, {Buchanan}, {Elitzur}, {Elvis}, {O'Dea}, \&
  {Robinson}}]{Gallimore2010}
{Gallimore}, J.~F., {Yzaguirre}, A., {Jakoboski}, J., {et~al.} 2010, \apjs,
  187, 172, \dodoi{10.1088/0067-0049/187/1/172}

\bibitem[{{Gon{\c{c}}alves} {et~al.}(2010){Gon{\c{c}}alves}, {Basu-Zych},
  {Overzier}, {Martin}, {Law}, {Schiminovich}, {Wyder}, {Mallery}, {Rich}, \&
  {Heckman}}]{Goncalves2010}
{Gon{\c{c}}alves}, T.~S., {Basu-Zych}, A., {Overzier}, R., {et~al.} 2010, \apj,
  724, 1373, \dodoi{10.1088/0004-637X/724/2/1373}

\bibitem[{{Goold} {et~al.}(2023){Goold}, {Seth}, {Molina}, {Ohlson}, {Runnoe},
  {Boeker}, {Davis}, {Dumont}, {Eracleous}, {Fern{\'a}ndez-Ontiveros}, {Gallo},
  {Goulding}, {Greene}, {Ho}, {Markoff}, {Neumayer}, {Plotkin}, {Prieto},
  {Satyapal}, {Van De Ven}, {Walsh}, {Yuan}, {Feldmeier-Krause},
  {G{\"u}ltekin}, {Hoenig}, {Kirkpatrick}, {L{\"u}tzgendorf}, {Reines},
  {Strader}, {Trump}, \& {Voggel}}]{Goold2023}
{Goold}, K., {Seth}, A., {Molina}, M., {et~al.} 2023, arXiv e-prints,
  arXiv:2307.01252, \dodoi{10.48550/arXiv.2307.01252}

\bibitem[{{Ho} \& {Keto}(2007)}]{Ho2007}
{Ho}, L.~C., \& {Keto}, E. 2007, \apj, 658, 314, \dodoi{10.1086/511260}

\bibitem[{{Kennicutt} \& {Evans}(2012)}]{Kennicutt2012}
{Kennicutt}, R.~C., \& {Evans}, N.~J. 2012, \araa, 50, 531,
  \dodoi{10.1146/annurev-astro-081811-125610}

\bibitem[{{Kroupa}(2001)}]{Kroupa2001}
{Kroupa}, P. 2001, \mnras, 322, 231, \dodoi{10.1046/j.1365-8711.2001.04022.x}

\bibitem[{{Lai} {et~al.}(2022){Lai}, {Armus}, {U}, {D{\'\i}az-Santos},
  {Larson}, {Evans}, {Malkan}, {Appleton}, {Rich}, {M{\"u}ller-S{\'a}nchez},
  {Inami}, {Bohn}, {McKinney}, {Finnerty}, {Law}, {Linden}, {Medling},
  {Privon}, {Song}, {Stierwalt}, {van der Werf}, {Barcos-Mu{\~n}oz}, {Smith},
  {Togi}, {Aalto}, {B{\"o}ker}, {Charmandaris}, {Howell}, {Iwasawa}, {Kemper},
  {Mazzarella}, {Murphy}, {Brown}, {Hayward}, {Marshall}, {Sanders}, \&
  {Surace}}]{Lai2022}
{Lai}, T. S.~Y., {Armus}, L., {U}, V., {et~al.} 2022, \apjl, 941, L36,
  \dodoi{10.3847/2041-8213/ac9ebf}

\bibitem[{{Lai} {et~al.}(2023){Lai}, {Armus}, {Bianchin}, {Diaz-Santos},
  {Linden}, {Privon}, {Inami}, {U}, {Bohn}, {Evans}, {Larson}, {Hensley},
  {Smith}, {Malkan}, {Song}, {Stierwalt}, {van der Werf}, {McKinney}, {Aalto},
  {Buiten}, {Rich}, {Charmandaris}, {Appleton}, {Barcos-Munoz}, {Boker},
  {Finnerty}, {Kader}, {Law}, {Brown}, {Hayward}, {Howell}, {Iwasawa},
  {Kemper}, {Marshall}, {Mazzarella}, {Muller-Sanchez}, {Murphy}, {Sanders}, \&
  {Surace}}]{Lai2023}
{Lai}, T. S.~Y., {Armus}, L., {Bianchin}, M., {et~al.} 2023, arXiv e-prints,
  arXiv:2307.15169, \dodoi{10.48550/arXiv.2307.15169}

\bibitem[{{Lang} {et~al.}(2019){Lang}, {Schinnerer}, {Smail},
  {Dudzevi{\v{c}}i{\={u}}t{\.{e}}}, {Swinbank}, {Liu}, {Leslie}, {Almaini},
  {An}, {Bertoldi}, {Blain}, {Chapman}, {Chen}, {Conselice}, {Cooke}, {Coppin},
  {Dunlop}, {Farrah}, {Fudamoto}, {Geach}, {Gullberg}, {Harrington}, {Hodge},
  {Ivison}, {Jim{\'e}nez-Andrade}, {Magnelli}, {Micha{\l}owski}, {Oesch},
  {Scott}, {Simpson}, {Smol{\v{c}}i{\'c}}, {Stach}, {Thomson}, {Toft},
  {Vardoulaki}, {Wardlow}, {Weiss}, \& {van der Werf}}]{Lang2019}
{Lang}, P., {Schinnerer}, E., {Smail}, I., {et~al.} 2019, \apj, 879, 54,
  \dodoi{10.3847/1538-4357/ab1f77}

\bibitem[{{Le Floc'h} {et~al.}(2001){Le Floc'h}, {Mirabel}, {Laurent},
  {Charmandaris}, {Gallais}, {Sauvage}, {Vigroux}, \& {Cesarsky}}]{LeFloch2001}
{Le Floc'h}, E., {Mirabel}, I.~F., {Laurent}, O., {et~al.} 2001, \aap, 367,
  487, \dodoi{10.1051/0004-6361:20000569}

\bibitem[{{Madau} \& {Dickinson}(2014)}]{Madau2014}
{Madau}, P., \& {Dickinson}, M. 2014, \araa, 52, 415,
  \dodoi{10.1146/annurev-astro-081811-125615}

\bibitem[{{Men{\'e}ndez-Delmestre} {et~al.}(2009){Men{\'e}ndez-Delmestre},
  {Blain}, {Smail}, {Alexander}, {Chapman}, {Armus}, {Frayer}, {Ivison}, \&
  {Teplitz}}]{MenendezDelmestre2009}
{Men{\'e}ndez-Delmestre}, K., {Blain}, A.~W., {Smail}, I., {et~al.} 2009, \apj,
  699, 667, \dodoi{10.1088/0004-637X/699/1/667}

\bibitem[{{Murphy} {et~al.}(2011){Murphy}, {Condon}, {Schinnerer}, {Kennicutt},
  {Calzetti}, {Armus}, {Helou}, {Turner}, {Aniano}, {Beir{\~a}o}, {Bolatto},
  {Brandl}, {Croxall}, {Dale}, {Donovan Meyer}, {Draine}, {Engelbracht},
  {Hunt}, {Hao}, {Koda}, {Roussel}, {Skibba}, \& {Smith}}]{Murphy2011}
{Murphy}, E.~J., {Condon}, J.~J., {Schinnerer}, E., {et~al.} 2011, \apj, 737,
  67, \dodoi{10.1088/0004-637X/737/2/67}

\bibitem[{{Nesvadba} {et~al.}(2010){Nesvadba}, {Boulanger}, {Salom{\'e}},
  {Guillard}, {Lehnert}, {Ogle}, {Appleton}, {Falgarone}, \& {Pineau Des
  Forets}}]{Nesvadba2010}
{Nesvadba}, N.~P.~H., {Boulanger}, F., {Salom{\'e}}, P., {et~al.} 2010, \aap,
  521, A65, \dodoi{10.1051/0004-6361/200913333}

\bibitem[{{Pope} {et~al.}(2008){Pope}, {Chary}, {Alexander}, {Armus},
  {Dickinson}, {Elbaz}, {Frayer}, {Scott}, \& {Teplitz}}]{Pope2008}
{Pope}, A., {Chary}, R.-R., {Alexander}, D.~M., {et~al.} 2008, \apj, 675, 1171,
  \dodoi{10.1086/527030}

\bibitem[{{Rich} {et~al.}(2023){Rich}, {Aalto}, {Evans}, {Charmandaris},
  {Privon}, {Lai}, {Inami}, {Linden}, {Armus}, {Diaz-Santos}, {Appleton},
  {Barcos-Mu{\~n}oz}, {B{\"o}ker}, {Larson}, {Law}, {Malkan}, {Medling},
  {Song}, {U}, {van der Werf}, {Bohn}, {Brown}, {Finnerty}, {Hayward},
  {Howell}, {Iwasawa}, {Kemper}, {Marshall}, {Mazzarella}, {McKinney},
  {Muller-Sanchez}, {Murphy}, {Sanders}, {Soifer}, {Stierwalt}, \&
  {Surace}}]{Rich2023}
{Rich}, J., {Aalto}, S., {Evans}, A.~S., {et~al.} 2023, \apjl, 944, L50,
  \dodoi{10.3847/2041-8213/acb2b8}

\bibitem[{{Riechers} {et~al.}(2014){Riechers}, {Pope}, {Daddi}, {Armus},
  {Carilli}, {Walter}, {Hodge}, {Chary}, {Morrison}, {Dickinson},
  {Dannerbauer}, \& {Elbaz}}]{Riechers2014}
{Riechers}, D.~A., {Pope}, A., {Daddi}, E., {et~al.} 2014, \apj, 786, 31,
  \dodoi{10.1088/0004-637X/786/1/31}

\bibitem[{{Rigby} {et~al.}(2008){Rigby}, {Marcillac}, {Egami}, {Rieke},
  {Richard}, {Kneib}, {Fadda}, {Willmer}, {Borys}, {van der Werf},
  {P{\'e}rez-Gonz{\'a}lez}, {Knudsen}, \& {Papovich}}]{Rigby2008}
{Rigby}, J.~R., {Marcillac}, D., {Egami}, E., {et~al.} 2008, \apj, 675, 262,
  \dodoi{10.1086/525273}

\bibitem[{{Sajina} {et~al.}(2012){Sajina}, {Yan}, {Fadda}, {Dasyra}, \&
  {Huynh}}]{Sajina2012}
{Sajina}, A., {Yan}, L., {Fadda}, D., {Dasyra}, K., \& {Huynh}, M. 2012, \apj,
  757, 13, \dodoi{10.1088/0004-637X/757/1/13}

\bibitem[{{Sales} {et~al.}(2010){Sales}, {Pastoriza}, \& {Riffel}}]{Sales2010}
{Sales}, D.~A., {Pastoriza}, M.~G., \& {Riffel}, R. 2010, \apj, 725, 605,
  \dodoi{10.1088/0004-637X/725/1/605}

\bibitem[{{Smercina} {et~al.}(2018){Smercina}, {Smith}, {Dale}, {French},
  {Croxall}, {Zhukovska}, {Togi}, {Bell}, {Crocker}, {Draine}, {Jarrett},
  {Tremonti}, {Yang}, \& {Zabludoff}}]{Smercina2018}
{Smercina}, A., {Smith}, J.~D.~T., {Dale}, D.~A., {et~al.} 2018, \apj, 855, 51,
  \dodoi{10.3847/1538-4357/aaafcd}

\bibitem[{{Spilker} {et~al.}(2023){Spilker}, {Phadke}, {Aravena}, {Archipley},
  {Bayliss}, {Birkin}, {B{\'e}thermin}, {Burgoyne}, {Cathey}, {Chapman},
  {Dahle}, {Gonzalez}, {Gururajan}, {Hayward}, {Hezaveh}, {Hill}, {Hutchison},
  {Kim}, {Kim}, {Law}, {Legin}, {Malkan}, {Marrone}, {Murphy}, {Narayanan},
  {Navarre}, {Olivier}, {Rich}, {Rigby}, {Reuter}, {Rhoads}, {Sharon}, {Smith},
  {Solimano}, {Sulzenauer}, {Vieira}, {Vizgan}, {Wei{\ss}}, \&
  {Whitaker}}]{Spilker2023}
{Spilker}, J.~S., {Phadke}, K.~A., {Aravena}, M., {et~al.} 2023, \nat, 618,
  708, \dodoi{10.1038/s41586-023-05998-6}

\bibitem[{{Stone} {et~al.}(2022){Stone}, {Pope}, {McKinney}, {Armus},
  {D{\'\i}az-Santos}, {Inami}, {Kirkpatrick}, \& {Stierwalt}}]{Stone2022}
{Stone}, M., {Pope}, A., {McKinney}, J., {et~al.} 2022, \apj, 934, 27,
  \dodoi{10.3847/1538-4357/ac778b}

\bibitem[{{Storey} \& {Hummer}(1995)}]{Storey1995}
{Storey}, P.~J., \& {Hummer}, D.~G. 1995, \mnras, 272, 41,
  \dodoi{10.1093/mnras/272.1.41}

\bibitem[{{U} {et~al.}(2022){U}, {Lai}, {Bianchin}, {Remigio}, {Armus},
  {Larson}, {D{\'\i}az-Santos}, {Evans}, {Stierwalt}, {Law}, {Malkan},
  {Linden}, {Song}, {van der Werf}, {Gao}, {Privon}, {Medling},
  {Barcos-Mu{\~n}oz}, {Hayward}, {Inami}, {Rich}, {Aalto}, {Appleton}, {Bohn},
  {B{\"o}ker}, {Brown}, {Charmandaris}, {Finnerty}, {Howell}, {Iwasawa},
  {Kemper}, {Marshall}, {Mazzarella}, {McKinney}, {Muller-Sanchez}, {Murphy},
  {Sanders}, \& {Surace}}]{U2022}
{U}, V., {Lai}, T., {Bianchin}, M., {et~al.} 2022, \apjl, 940, L5,
  \dodoi{10.3847/2041-8213/ac961c}

\bibitem[{{U} {et~al.}(2023){U}, {Lai}, {Bianchin}, {Remigio}, {Armus},
  {Larson}, {Diaz-Santos}, {Evans}, {Stierwalt}, {Law}, \& {Goals
  Collaboration}}]{U2023}
{U}, V., {Lai}, S.-Y.~T., {Bianchin}, M., {et~al.} 2023, in American
  Astronomical Society Meeting Abstracts, Vol.~55, American Astronomical
  Society Meeting Abstracts, 418.07

\bibitem[{{Verma} {et~al.}(2003){Verma}, {Lutz}, {Sturm}, {Sternberg},
  {Genzel}, \& {Vacca}}]{Verma2003}
{Verma}, A., {Lutz}, D., {Sturm}, E., {et~al.} 2003, \aap, 403, 829,
  \dodoi{10.1051/0004-6361:20030408}

\bibitem[{{Wright} {et~al.}(2023){Wright}, {Rieke}, {Glasse}, {Ressler},
  {Garc{\'\i}a Mar{\'\i}n}, {Aguilar}, {Alberts}, {{\'A}lvarez-M{\'a}rquez},
  {Argyriou}, {Banks}, {Baudoz}, {Boccaletti}, {Bouchet}, {Bouwman}, {Brandl},
  {Breda}, {Bright}, {Cale}, {Colina}, {Cossou}, {Coulais}, {Cracraft}, {De
  Meester}, {Dicken}, {Engesser}, {Etxaluze}, {Fox}, {Friedman}, {Fu},
  {Gasman}, {G{\'a}sp{\'a}r}, {Gastaud}, {Geers}, {Glauser}, {Gordon},
  {Greene}, {Greve}, {Grundy}, {G{\"u}del}, {Guillard}, {Haderlein},
  {Hashimoto}, {Henning}, {Hines}, {Holler}, {Detre}, {Jahromi}, {James},
  {Jones}, {Justtanont}, {Kavanagh}, {Kendrew}, {Klaassen}, {Krause},
  {Labiano}, {Lagage}, {Lambros}, {Larson}, {Law}, {Lee}, {Libralato}, {Lorenzo
  Alverez}, {Meixner}, {Morrison}, {Mueller}, {Murray}, {Mycroft}, {Myers},
  {Nayak}, {Naylor}, {Nickson}, {Noriega-Crespo}, {{\"O}stlin}, {O'Sullivan},
  {Ottens}, {Patapis}, {Penanen}, {Pietraszkiewicz}, {Ray}, {Regan},
  {Roteliuk}, {Royer}, {Samara-Ratna}, {Samuelson}, {Sargent}, {Scheithauer},
  {Schneider}, {Schreiber}, {Shaughnessy}, {Sheehan}, {Shivaei}, {Sloan},
  {Tamas}, {Teague}, {Temim}, {Tikkanen}, {Tustain}, {van Dishoeck},
  {Vandenbussche}, {Weilert}, {Whitehouse}, \& {Wolff}}]{Wright2023}
{Wright}, G.~S., {Rieke}, G.~H., {Glasse}, A., {et~al.} 2023, \pasp, 135,
  048003, \dodoi{10.1088/1538-3873/acbe66}

\bibitem[{{Yan} {et~al.}(2007){Yan}, {Sajina}, {Fadda}, {Choi}, {Armus},
  {Helou}, {Teplitz}, {Frayer}, \& {Surace}}]{Yan2007}
{Yan}, L., {Sajina}, A., {Fadda}, D., {et~al.} 2007, \apj, 658, 778,
  \dodoi{10.1086/511516}

\bibitem[{{Zamojski} {et~al.}(2011){Zamojski}, {Yan}, {Dasyra}, {Sajina},
  {Surace}, {Heckman}, \& {Helou}}]{Zamojski2011}
{Zamojski}, M., {Yan}, L., {Dasyra}, K., {et~al.} 2011, \apj, 730, 125,
  \dodoi{10.1088/0004-637X/730/2/125}

\bibitem[{{Zhuang} {et~al.}(2019){Zhuang}, {Ho}, \& {Shangguan}}]{Zhuang2019}
{Zhuang}, M.-Y., {Ho}, L.~C., \& {Shangguan}, J. 2019, \apj, 873, 103,
  \dodoi{10.3847/1538-4357/ab0650}

\end{thebibliography}
\bibliographystyle{aasjournal}

\end{document}